\documentclass{article}

\usepackage{arxiv}

\usepackage{amsmath}
\usepackage{amsthm}
\usepackage{amsfonts}
\usepackage[numbered]{bookmark}
\usepackage{float}
\usepackage{graphicx}
\usepackage{caption}
\usepackage{subcaption}
\usepackage{textcomp}
\usepackage{lscape}
\usepackage{enumitem}
\usepackage{booktabs}
\usepackage{textpos}
\usepackage{chngcntr}
\usepackage{url}
\usepackage{xcolor,colortbl}
\usepackage{hyperref}
\usepackage{array,booktabs,arydshln}
\usepackage{verbatim}
\usepackage{calrsfs}

\usepackage{url}
\usepackage{xcolor} 
\usepackage{hyperref}
\usepackage{array,booktabs,arydshln}
\usepackage{soul}
\usepackage{afterpage}

\newcommand{\E}{\mathbb{E}}
\newcommand{\Q}{\mathbb{Q}}

\newcommand{\mc}{\mathrm{mc}}

\newcommand{\SIMM}{\mathrm{SIMM}}
\newcommand{\IM}{\mathrm{IM}}
\newcommand{\DeltaMargin}{\mathrm{DeltaMargin}}
\newcommand{\VegaMargin}{\mathrm{VegaMargin}}
\newcommand{\CurvatureMargin}{\mathrm{CurvatureMargin}}
\newcommand{\BaseCorrMargin}{\mathrm{BaseCorrMargin}}

\newcommand\VRule[1][\arrayrulewidth]{\vrule width #1}
\newcommand{\norm}[1]{\left\lVert#1\right\rVert}

\definecolor{myGray}{rgb}{0.9,0.9,0.9}

\newtheorem*{thm*}{Theorem}
\theoremstyle{remark}
\newtheorem{rem}{Remark}[section]
\newtheorem*{rem*}{Remark}

\definecolor{DarkBlue2}{rgb}{0,0.08,0.50} 
\hypersetup{
    colorlinks = true, 
    linkcolor = DarkBlue2, 
    citecolor = DarkBlue2, 
    urlcolor = blue, 
    anchorcolor = red
}

\hypersetup{
pdftitle={Efficient ISDA Initial Margin Calculations Using Least Squares Monte-Carlo},
pdfsubject={q-fin.CP},
pdfauthor={Asif Lakhany, Amber Zhang},
pdfkeywords={ISDA Standard Initial Margin Model, Nested Monte Carlo Simulation, Least-Squares Monte-Carlo, Sensitivity Estimation}
}

\title{Efficient ISDA Initial Margin Calculations Using Least Squares Monte-Carlo}
\date{\today}

\author{
    \href{https://orcid.org/0000-0002-3509-230X}{\includegraphics[bb = 0 0 128 128, scale=0.06]{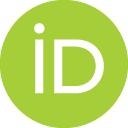}\hspace{1mm}Asif Lakhany} \\
    FinTech Consultant\\
    Markham, Ontario, Canada \\
    \texttt{Asif.Lakhany@riskconsortium.ai} \\
    \And
    \href{https://orcid.org/0000-0003-0247-8107}{\includegraphics[bb = 0 0 128 128, scale=0.06]{./orcid.png}\hspace{1mm}Amber Zhang} \\
    SS\&C Algorithmics \\
    Toronto, Canada \\
    \texttt{Amber.Zhang@sscinc.com} \\
}

\begin{document}
\maketitle

\begin{abstract}
              Non-cleared bilateral OTC derivatives between two financial firms or systemically 
              important non-financial entities are subject to regulations that require the
              posting of initial and variation margin. The ISDA standard approach (SIMM)
              provides a way for computing the initial margin. It involves computing
              sensitivities of the contracts with respect to several market factors. In this paper, the authors extend the well known LSMC technique
              to efficiently estimate the sensitivities required
              in the ISDA SIMM methodology.
\end{abstract}


\maketitle

\section{Introduction} 
\label{sec:Introduction} 
During the financial crisis of 2008, the lack of transparency in the OTC derivative market made it difficult 
for governments and regulators to assess the amount of risk that had been built up in the system due to this type of trading. 
To address this shortfall, as well as to reduce the counterparty credit risk inherit in bilateral OTC derivatives, post-crisis, 
national governments initiated reforms to encourage derivatives to be traded in exchanges and cleared through central clearing counterparties (CCPs) whenever possible.

In 2011 the Basel Committee on Banking Supervision (BCBS) and the Board of the International Organization of 
Securities Commission (IOSCO) formed the Working Group on Margin Requirements (WGMR) with the aim to develop 
international standards for the new margin reforms. In September 2013, the WGMR issued a final margin policy framework for non-cleared 
bilateral derivatives ( \cite{basel_margin} ). The margin requirements have a twofold benefits. In the event of default of one of the counterparties, 
the margin absorbs the losses to the surviving counterparty.  Additionally, from the regulators' perspective, these regulations are 
designed to make the bilateral agreements less appealing compared to its cleared counterparts. As a result, these regulations 
would encourage market players to trade their transactions more inside CCP and therefore, force more transparency into the  
financial industry. However, at the same time, it is recognized that not all contracts can be done through CCP. 

Computing Initial Margin (IM) is important. For example, it is necessary for pricing a credit valuation adjustment (CVA) or potential future exposure (PFE). Since IM is tied to the models used, this can produce inconsistent results among counterparties in a bilateral agreement creating a possibility for dispute. Based on this, the International Swap and Derivatives Association (ISDA) came up with a standard method for computing the IM called the Standard Initial Margin Model (SIMM). SIMM is rooted on the Basel Committee’s sensitivity-based approach (\cite{ISDA}). 

Meanwhile, estimating the sensitivities required for the calculation of the IM under the SIMM methodology within a nested Monte Carlo framework could be prohibitively expensive. This is the consequence of the fact that a portion of banking book may contain exotic instruments which may require a Monte Carlo (MC) method for pricing and sensitivity estimation. As such, this will require large number of Monte Carlo paths for sufficiently accurate estimation of the sensitivities for IM calculations. Just to visualize the complexity of the problem, we give an example here. A typical banking book may contain upwards of $150,000$ positions. Even if we assume that only 10\% of these instruments are exotics and require a MC method, that number comes out to be $15,000$ instruments. Typically, 5000 outer multi-step Monte Carlo scenarios are used with $100$ time steps. Even if we assume a modest 10 sensitivities per instruments on average (which is rather a very low number for cross currency exotics) and a central finite difference approach for sensitivity estimation, we end up with $15,000*5000*100*10*2$ price estimates. Since sensitivities are hard to capture, we expect to at least need $10,000$ paths on average to accurately estimate the required sensitivities. The total number of the Monte Carlo paths that will be processed, comes to be $15,000*5,000*100*10*2*10,000= 1E+15$. This is an astronomical number and can put immense strain on any risk management system, no matter how efficiently designed.  In this paper, we propose to use the Least Squares Monte Carlo (LSMC) method 
(c.f. \cite{barrie_hibbert11, lsmc}) for the estimation of sensitivities of exotic instruments through the Monte Carlo method. We show that using this approach can greatly speed up the ISDA SIMM calculations, while achieving the same level of accuracy.\footnote{Later in the paper, we also propose filtering scenarios that can further reduce the computation cost. In addition to this, techniques like adjoint differentiation, discussed later in the paper, can work in tandem with the LSMC method which can also reduce the computational cost for estimating sensitivities.}

The breakdown of this paper as follows. In Section \ref{sec:LSMC} we provide a brief introduction on the LSMC based approach. Then we walk through 
the list of sensitivities required in SIMM calculations in Section \ref{sec:ISDASIMM}.
Our main
results are produced in Section \ref{sec:MAIN}. The accuracy tests of the LSMC method in terms of the SIMM estimation at a 
portfolio level are presented in Section \ref{sec:SIMMEstimation}. We wrap up with concluding remarks in Section \ref{sec:FINAL}.

\section{LSMC in a Nutshell}
\label{sec:LSMC}
The LSMC method works under a nested Monte Carlo simulation. In this case, we have a Monte Carlo scenarios set over a discrete set of time points such as in (\ref{eq:discrete}). At each point in this time set we have $n$ scenarios on market factors and for each such scenario we have to price a set of financial instruments. Some of these financial instruments could be exotic and require a Monte Carlo method for pricing, thereby resulting in the term ``Nested Monte Carlo''. Large number of paths may be necessary to estimate the price of exotic instruments. In what follows, this setting is referred to as a Full Monte Carlo (FMC) method. In contrast, the LSMC method can produce the same quality of price estimates with a far fewer number of paths. The basic idea lies in the ability of building a smooth conditional expectation function, using the cross sectional information available at any time, through regression. The cross sectional information is the set of explanatory variables and the crude price estimates. When applied to estimation of sensitivities, the same information would be the set of explanatory variables and crude estimates of the sensitivities of interest.

\subsection{Detailed Description of Least Squares Monte Carlo}
Least Squares Monte-Carlo (LSMC) was 
developed 
in \cite{barrie_hibbert11}. It was subsequently analyzed
in \cite{lsmc}. Suppose that $n$ Monte Carlo scenarios have been generated over $K$ times points, $\mathcal{T} \equiv \lbrace t_j, j=1,2,\ldots,K \rbrace$, such that:
\begin{equation}
	t_1 < t_2 <\cdots < t_{K-1} < t_K
\label{eq:discrete}
\end{equation}

For any time $t \in \mathcal{T}$, our cross sectional information is represented by
$n$ vectors $\mathbf{x}_{t,i}, i=1,2,\ldots,n$. This information is build using the
$n$ outer scenarios at time $t$ as well as all the information absorbed until time $t$.\footnote{The set of this additional information could possibly include  
average so far, redemption level, barrier breach, etc. The vectors
$\mathbf{x}_{t,i}, i=1,2,\ldots,n$ form the set of explanatory variables at time $t$ for the regression model.
}
Suppose now, that we are interested in estimating the time $t$ price of an exotic
instrument maturing at time $T$. If $f(\mathbf{x}_{t,\, i})$ denotes the
discounted payoff, then the time $t$ price is the conditional expectation

\begin{equation} 
    \label{eq:price} 
    y_i := \E^\Q \Big[ f(\mathbf{x}_{t,\, i}) \, \Big|\, \mathcal{F}_t \Big] .
\end{equation}

One can estimate the integral on the right-hand-side of equation
\eqref{eq:price} using a Monte Carlo (MC) method. Doing so, will generate a sequence of MC
estimates $y_{\mc,\, i}$, $i = 1,\, 2,\, \dots,\, n$ at each time
$t \in \mathcal{T}$, where $\mathcal{T}$ represents a set of future time points. Specifically, 
\begin{equation}
    \label{eq:mc_price}
    y_{\mc,\, i} = \frac{1}{p} \sum_{k = 1}^{p}{ f_{k}(\mathbf{x}_{t,\, i})}, 
\end{equation}
where $\mathbf{x}_{t,\, i}$ is generated under risk neutral measure, and
$f_k(\cdot)$ is the value of the discounted payoff function on the $k^{th}$ 
path.\footnote{In this nested Monte-Carlo framework which is used to estimate the conditional expectation, 
we refer to the inner loop samples as ``paths'' and the outer loop samples as ``scenarios".}
The accuracy of our estimate in (\ref{eq:mc_price}) depends on the number of paths $p$ used.
For computing sensitivities, this number could be substantially high - depending
on the sensitivity type and the instrument type. What LSMC offers is the reduction in the number of paths used to achieve a reasonably
accurate estimate of the sensitivities. The reduction could be an order of one or two in magnitude.
Instead of estimating the future prices using a large enough $p$, we obtain crude
estimation of the same by significantly reducing the number of paths $p$.  These prices 
form the corresponding set of response
variables for our regression model. We can write the relationship as
\begin{equation}
    \label{eq:mc_approx}
    y_{\mc,\, i} = \sum_{j=0}^m \beta_j\, b_j(\mathbf{x}_{t,\, i}) + \xi_{i} 
\end{equation}
where $\{\beta_{j}\}_{j=0}^m$ are coefficients of expansion and $\{b_j(\mathbf{x}_{t,\, i}\}_{j=0}^m$ are basis functions
of choice. For example, in \cite{lsmc} orthogonal Forsythe polynomials are used as basis
functions. We can write the regression model (\ref{eq:mc_approx}) in a matrix notation as
\begin{align} 
    \label{eq:reg_mc} 
    \mathbf{y}_{\mc} &= \mathbf{X} \boldsymbol{\beta} +
    \boldsymbol{\xi}.
\end{align} 
Then, the coefficient estimator is obtained as $\hat{\boldsymbol{\beta}} =
(\mathbf{X^TX})^{-1} \mathbf{X^Ty_{\mc}}$, and the LSMC estimator is $\hat{\bf{y}}_{\mc} = \bf{H}
\bf{y}_{\mc}$ where 
$\bf{H} = \bf{X}(\mathbf{X^TX})^{-1} \mathbf{X^T}.$

Perhaps the best way to cover the algorithm in this paper is by reproducing an example
from our previous paper \cite{lsmc} where  an Arithmetic
Asian Option example is used to check the fit by using polynomials.
Consider an Arithmetic Asian Put Option that matures at time $T$ with
fixings $\{t_i\}_{i = 1}^{s}$ and weights $\{w_i\}_{i = 1}^{s}$.  The payoff at
maturity takes the form 
\begin{equation}
    \text{payoff} = (K  - \sum_{i = 1}^{s} { w_i S_{t_i} })^{+},\,
    \text{ $0 \leq t_i \leq T\quad \forall i$ }, 
    \label{eq:asian_option}
\end{equation}
where $S_{t_{i}}$ is the price of the underlying equity at time $t_i$. If we
 assume that $S$ follows a Geometric Brownian Motion with constant drift and volatility in
 both the outer scenarios and the inner Monte Carlo paths.\footnote{The drift in the outer loop is set at 0.1,
while the drift in the inner loop is set at 0.05.} Let the observation time be
$t_k$ for $1 < k \leq s$.  We use the orthogonal Forsythe polynomials
\cite{forsythe_poly} for basis functions using $B_{t_k} = \sum_{i}^k w_i  S_i$
as the explanatory variable in the regression model. Following steps are
required to obtain an LSMC estimate:
\begin{itemize}
    \item Standardize the explanatory variables in the interval $[-1,1]$:
        \begin{equation*}
            B_{t_k,\, j}^{*} := \frac{2 B_{t_k,\, j} - \max(\mathbf{B}_{t_k}) -
            \min(\mathbf{B}_{t_k})}{\max(\mathbf{B}_{t_k}) - \min(\mathbf{B}_{t_k})} 
        \end{equation*}
    \item Obtain regression matrix $\mathbf{X}$ using Forsythe polynomial
        expansion \nocite{forsythe_poly}
    \item Estimate the coefficients $\boldsymbol{\beta}$ in model \eqref{eq:mc_approx}
    \item Use the coefficient estimates to obtain LSMC prices
\end{itemize}
We run a numerical experiment with 1, 10, 30, 50, 100 and 10,000 inner paths
to compute and regress $\mathbf{Y}_{\mc}$ against $\mathbf{X}$ to get
$\hat{\mathbf{Y}}_{\mathrm{\mc}}$. We compare LSMC estimates to MC prices obtained using
$131,072$ Sobol paths. The results are summarized in Figure \ref{fig:fit}. From this figure, we infer that we can obtain a very accurate price estimates using only 10 paths per outer scenario. 
\afterpage{%
\begin{landscape}
\begin{figure}
    \centering
    \captionsetup{justification=centering}
    \caption{MC vs. LSMC Price: \\ First time step fit using polynomial of
    degree five}
    \includegraphics[width=1.3\textwidth]{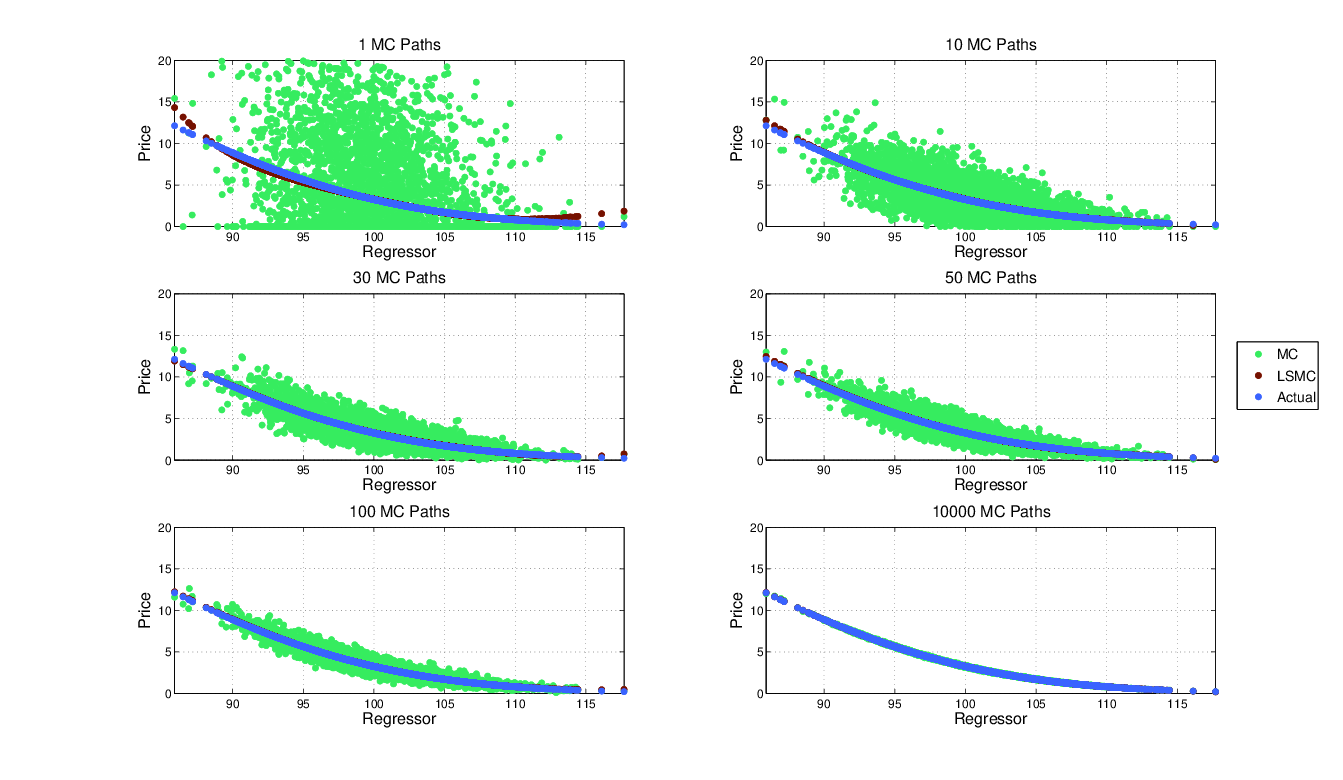}
    \label{fig:fit}
\end{figure}
\end{landscape}
}

\subsection{Thin Plate Splines}
In this paper we also extend the traditional least squares method, as in the seminal work of
Longstaff and Schwartz (c.f. \cite{longstaff}) using polynomial basis functions, to
a more general smoothing algorithm using splines. In particular,
we propose to use Thin Plate Splines (TPS). We show this technique to be very powerful, specifically
when working with instruments with complicated payoffs. In what follows, we introduce the TPS
for two dimensional case (such as the spot price and the one factor of Hull-White interest rate model).\footnote{For extensions to higher dimension, we refer to the excellent monograph by Grace Wahba (\cite{WahbaBook}).}
In what follows, our exposition is based on the article \cite{Kamp}.

Let $(x_i, y_i)$ be a given set of data points where estimated values $z_i$ are given. 
For a given $\lambda$, the TPS fit, $f(x_i,y_i)$, is obtained by minimizing the functional
\begin{equation}
\mathcal{E} = {{1}\over{n}} \sum_{i=1}^n \left( f(x_i,y_i) - z_i \right)^2 + \lambda \mathcal{J}
\label{eq:energy}
\end{equation}
where
\begin{equation*}
\mathcal{J} = \iint_{{\mathbb R}^2} \left( \left({\partial^2 f}\over{\partial x^2}\right)^2 +
    2\left({\partial^2 f}\over{\partial x \partial y}\right)^2 + \left({\partial^2 f}\over{\partial y^2}\right)^2
    \right) dx dy.
\end{equation*}
In essence, the right hand side of Equation (\ref{eq:energy}) suggests that besides a regular
least squares fitting term, there is an additional term that minimizes the bending energy of the
TPS fit. In absence of this term, the TPS fit will pass through every single point of the data. Hence, for the LSMC application, we need
$\lambda > 0$ so as not to collocate the noisy Monte Carlo estimate. 

The unique solution to the problem (\ref{eq:energy}) can be given by the following expansion (c.f. \cite{Duchon} and \cite{Meinguet} ):
\begin{equation}
f(x_i, y_i) = \sum_{i=1}^n a_j A_{i,j} + \sum_{j=1}^3 b_j B_{i,j}	
\label{eq:tpsexp}
\end{equation}
where
\begin{align}
A_{i,j} &= \norm{(x_i,y_i) - (x_j,y_j)}^2 \log { \left(	\norm{(x_i,y_i) - (x_j,y_j)}^2 \right) }
\nonumber\\
B_{i,\cdot} &= \left[ 1\ x_i\ y_i \right].
\label{eq:tpsAB}
\end{align}
The coefficients, $\bold a \equiv \left[ a_1 \ldots a_n \right]^T$ and $\bold b \equiv \left[b_1\ b_2\ b_3\right]^T$ that minimize the energy in Equation (\ref{eq:energy}) are given by (\cite{Kamp}, \cite{Wahba}) :

\begin{align}
	\bold a &= \bold Q^T_2 \left(\bold Q_2^T \bold A \bold Q_2 + n\lambda \bold I\right)^{-1} \bold Q_2^T \bold f := \hat {\bold A} \bold f \left( \rm{say}\right)\nonumber\\
	\bold b &= \bold R_1^{-1} \bold Q_1^T \left( \bold f - \bold A \bold a\right) = \bold R_1^{-1} \bold Q_1^T \left( \bold I  - \bold A \hat{\bold A} \right) \bold f := \hat{\bold B} \bold f \left(\rm{say}\right)
\label{eq:tpscoeff}
\end{align}
where $\bold A$ and $\bold B$ are the matrices obtained from the coefficients in Equation (\ref{eq:tpsAB}), and the matrices $\bold Q_1, \bold Q_2$ and $\bold R_1$ appear in the $QR$ factorization of the matrix $\bold B$:
\begin{align*}
	\bold B &= \bold Q\bold R = \left[ \bold Q_1 \vert \bold Q_2 \right] \left[ 
	\begin{aligned}
	    & \bold R_1\\ & \bold 0 	
	\end{aligned} \right] = \bold Q_1 \bold R_1\nonumber\\
	\bold f &= \left[ f(x_1,y_1) \ldots f(x_n,y_n)\right]^T.
\end{align*}
The important feature in Equation set (\ref{eq:tpscoeff}) that we shall exploit, in what follows, is the fact that all the expensive calculations can be done before the data $f$ enters the calculation.
 \vskip 0.1in
\subsection{Reporting performance gains from Least Squares Monte Carlo}
\label{subsec:perf_lsmc}
Now that we know where the gains from using the the Least Squares Monte Carlo is coming from, the very question that we have to address is how to report this performance gain. One obvious way would be to actually run several tests on different hardwares. This would be a prudent choice in a production pipeline. However, since this type of tests would significantly depend on how the LSMC was implemented in a software system, it is not a very useful metric for the current paper. As such, in what follows we leverage from the fact that the cost of processing a single LSMC path is identically same as the cost of processing a single FMC path, thereby, the difference in the two methods only lie in how many paths are required for an accurate enough estimation. This observation simplifies the problem. For instance, let us assume that the FMC requires $p$ number of paths to get a reasonably accurate estimate for the quantity in question (be it price or sensitivity). If, now, the LSMC is able to provide the same quality of estimation with $q << p$ paths, then based on the perfectly linear cost of the Monte Carlo method in the number of paths; the potential speedup is given by $p/q$. This is the preferred choice for reporting speed up due to LSMC in this paper.

At this stage, one may argue that the potential speedup may not, necessarily, be a very realistic estimate of the speedup. We agree that the actual speedup will be less than the potential speed up that we report in this paper; nevertheless, we argue that in absence of the availability of other information, such as the implementational details, this is still the best number we can produce. Let us strengthen this argument with a simple example. The TPS method, for instance, may benefit from some data filtering of cross sectional information. One can, for example, perform this at every simulation time point during the nested Monte Carlo simulation -- or simply execute this task when the scenarios are generated. Moreover, as noted above, some matrix manipulations can be pre-calculated. Last, but not the least, we need to acknowledge the fact that the cost of regression is amortized over the number of scenarios, $n$, for a cross section. Hence the overhead of regression for a single scenario for a given time step is quite negligible. This justifies our reporting of the potential speed up from using the LSMC method. On the other hand, data filtering can possibly make the actual speedup to be greater than the potential speedup.

\section{ISDA SIMM in a Nutshell}
\label{sec:ISDASIMM}
This paper applies the LSMC method on the computation of IM which depends on potential future exposure. As such, it is predominantly model dependent.
This allows for the possibility that each of the counterparty in a bilateral deal may come up with 
its own margin calculation - which may or may not agree with the other counterparty's calculation.
ISDA SIMM approach handles that by standardizing the IM calculation. The calculation
of ISDA SIMM is detailed in the document \cite{ISDA}. In this section, we
merely list the sensitivities to be computed as well as the way in which they are
grouped and computed. 
Each trade is assigned to one of the four product classes:
\begin{itemize}
\item{RatesFX} : Interest Rates and Foreign Exchange
\item{Credit}
\item{Equity}
\item{Commodity}
\end{itemize}

Meanwhile, for each of the above product classes, the following risk classes are defined:
\begin{itemize}
\item{Interest Rate}
\item{Credit (Qualifying)}
\item{Credit (Non-Qualifying})
\item{Equity}
\item{Commodity}
\item{FX}
\end{itemize}

The final SIMM value is calculated by aggregating individual product classes' SIMM values in the
following formula:
\begin{equation}
\SIMM = \SIMM_{RatesFX} + \SIMM_{Credit} + \SIMM_{Equity} + \SIMM_{Commodity}.
\label{eq:SIMM}
\end{equation}

Additionally, SIMM value for the individual product classes appearing on the right hand side of
equation (\ref{eq:SIMM}) is computed using the following formula:
\begin{equation}
\SIMM_{product} = \sqrt{ \sum_r \IM_r^2 + \sum_{r}\sum_{s\neq r} \psi_{rs}\IM_r\IM_s }
\label{eq:SIMMP}
\end{equation}
where $r$ and $s$ are risk classes, whose correlations $\psi_{rs}$ are tabulated in Section K of ISDA document.

The margin for each of the above six risk class is computed as the sum of four
margins given in the following formula:
\begin{equation}
\IM_r = \DeltaMargin_r + \VegaMargin_r + \CurvatureMargin_r + \BaseCorrMargin_r
\label{eq:IM}
\end{equation}
where the $\BaseCorrMargin$ is only present in the Credit Qualifying case.
The IMs computed from formula (\ref{eq:IM}) are plugged on the right hand side
of formula (\ref{eq:SIMMP}) which is then used in formula (\ref{eq:SIMM}) to
provide us an estimate of the standard initial margin. The missing pieces are
now the margins appearing on the right hand side of formula (\ref{eq:IM}). 
These are computed from a collection of formul\ae\ based on forward sensitivities.
Readers are suggested to refer to the ISDA document \cite{ISDA} for details of the calculations.

\section{Numerical Results For Sensitivity Estimation}\label{sec:MAIN}
In this section we shall apply the LSMC technique to the estimation of 1) Delta, 2) Vega and 3) Rho sensitivities 
that are required by the ISDA SIMM calculator. In essence, instead of applying the LSMC technique to prices, as described in the
previous section, we would apply it directly to smooth the sensitivities estimated using the reduced number of paths on any
cross section of time. At this point, some may argue that one can always apply the LSMC technique on two sets of values
(for instance, one obtained using the base value of a chosen riskfactor and the other obtained by bumping up the same riskfactor)
and estimate sensitivities using the finite difference method. 
Not only this method is not guaranteed to produce reliable estimates, but it would also not allow leveraging from the adjoint differentiation techniques described in the sequel which is considered a de facto standard of modern risk management systems.

We shall compare our results with a benchmark computation using one million paths. Of course, even with large number of paths, there still will be some Monte Carlo error, but we presume that it will be small enough for comparison purposes. Initial margins are computed at the counterparty level. The number of trades at this level
may range from a single trade to several thousands. As a rule of thumb we generally consider 20-25 percent
exotics versus 75-80 percent vanilla instruments for the sake of analysis. As such, our test
portfolio consists of two exotic instruments and six vanilla instruments. 
Table \ref{tab:instr} shows that the proportion of exotic instruments in our test portfolio is around 20\% in terms of dollar values. 
In order to ease our calculations, we have made a few simplifications. These simplifications do not undermine
the applicability of LSMC techniques for sensitivity estimation. 
These are enumerated below:
\begin{enumerate}[]
\item The window Up-and-Out Barrier is set up such that the simulation date is before 
the window start date. This is just to make sure that the Barrier option is still
alive at the simulation time.
\item All contributions to the initial margins are coming from a single risk class. 
\item The underlying instruments belong to the same equity bucket, resulting in simplification of the SIMM formula.
\end{enumerate}
The details are provided in Tables \ref{tab:instr} and \ref{tab:under}.
\begin{table}[h]
\small
\caption{Details of the test portfolio }
\centering
\begin{tabular}{!{\VRule} l !{\VRule} c !{\VRule} c !{\VRule} c !{\VRule}}
\hline
Type &	Maturity	& Underlying	 & Price  \\
\hline
European 1Y ATM Put	& 272 days	& MCK	& 31.21 USD\\
European 2Y ATM Call &	455 days	& ABC	& 17.33 USD\\
EQ Forward &	455 days	 & MCK	& 35.91 USD\\
EQ Forward &	455 days	& ABC &	40.87 USD\\
EQ Futures &	455 days	 & ABC	& 214.73 USD\\
EQ Futures &	455 days	 & MCK	& 140.17 USD\\
Arithmetic Avg. Asian Option	& 272 days& 	ABC	& 104.1 USD\\
Barrier Option &	272 days	& ABC	& 11.12 USD\\
\hline
\end{tabular}
\label{tab:instr}
\end{table} 

\begin{table}[h]
\small
\caption{Underlying equity instruments}
\centering
\begin{tabular}{!{\VRule} l !{\VRule} c !{\VRule} c !{\VRule} c !{\VRule}}
\hline
Underlying stock	& Company Name	& Industry	& Equity Bucket Number\\
\hline
MCK	& McKesson Corporation	& pharmaceuticals & 	5\\
ABC &	AmerisourceBergen Corp	& pharmaceuticals &	5\\
\hline
\end{tabular}
\label{tab:under}
\end{table} 

Our first test includes estimating the sensitivities of the Asian Option described in Tables \ref{tab:instr} and \ref{tab:under}. The risk factors of the Asian Option are underlying stock prices (S) and the discount curve. By assuming the stock prices follow the Geometric Brownian Motion (GBM) process and interest rates follow Hull-White one factor (HW1F) model, we simulate the whole movement process of the target Asian Option along the time. In order to examine the accuracy of the sensitivities estimated using the LSMC method, we select the FMC-based estimation results with one million paths generated by the BRODA Quasi Monte Carlo generator (c.f. \cite{BRODA}) as the benchmark.

There are 13 term nodes on the discount curve. In view of the fact, that the whole interest rate curve is generated by HW1F model, we can easily avoid the multicollinearity problem by using a single factor ($P$) to represent the explanatory variable for the interest rate part. This can easily be achieved by using the principal component analysis and retaining only a single component.
Based on the above observations, our final LSMC regression for the Asian Option is:\footnote{This is an expanded form of formula (\ref{eq:mc_approx}) for the purpose of illustration.}
\begin{equation}
\Delta^{Asian}_{i} = c_0 + c_1 * S_{i} + c_2 *P_{i} + c_3* S_{i}^{2} + c_4*P_{i}^{2} + c_5*S_{i}*P_{i} + \epsilon_{i}
\end{equation}
where, $i =  1, 2, ..., 5000$.

We examine the estimation accuracy of the LSMC method with the assistance of some statistical measurements. 
One of the most commonly used is Kolmogorov Smirnoff (KS) test. This  test is based on the ''D statistics'' 
and the null hypothesis is whether two given samples are from the same distributions with certain confidence 
level (usually taken as 95\%) (c.f. \cite{gof}). When the p-value of KS test is smaller than 5\%, we will reject the null hypothesis. 
Table \ref{tab:gof} shows the p-value results for the KS test with the null hypothesis that LSMC estimations 
with different numbers of inner paths and benchmark values are from the same distribution. For reader's convenience, 
we greyed out the results which do not pass the KS test at 95\% confidence level. Clearly,  when we select inner path 
number that is greater than 64, the LSMC estimation distributions are statistically consistent with the benchmark. If
now one assumes that a regular Full Monte Carlo (FMC) method requires 4096 paths\footnote{a popular choice for smooth payoff functions}, the potential speedup is 64 times, less the overhead. Of course, one can
be aggressive and take only 16 paths (based on the QQ plot), in which case the potential speedup is 256 times, without the overhead.
 
\begin{table}[h]
\small
\label{tab:gof}
\caption{Arithmetic Asian Option : Kolmogorov - Smirnov Test}
\centering
\begin{tabular}{!{\VRule} l !{\VRule} c !{\VRule} c !{\VRule} c !{\VRule} c !{\VRule} c !{\VRule} c !{\VRule}}
\hline
 \# of Paths & Delta 	&	 Vega 	&	  Rho 14d &		  Rho 1m 	&	  Rho 3m 	& 	  Rho 6m \\	
\hline
16	& \cellcolor{myGray} 0.02279	&	\cellcolor{myGray} 0.0092 &		\cellcolor{myGray} 0.0080	&	\cellcolor{myGray} 0.0069 &		\cellcolor{myGray} 0.0080	&	0.0518 \\
32	& \cellcolor{myGray} 0.0336	&	\cellcolor{myGray} 0.0336 &		\cellcolor{myGray} 0.0336	&	\cellcolor{myGray} 0.0336 &		\cellcolor{myGray} 0.0296	& 	0.1203 \\
64	& 0.1478	&	0.0735	&	0.1478	&	0.1405	&	0.1478	&	0.2612 \\
128	& 0.7509	&	0.3662	&	0.6960 &		0.7145	&	0.7145	&	0.8665 \\
256	& 0.9819	&	0.5121	&	0.9648 &		0.9771	&	0.9860	&	0.9996 \\
\hline
\end{tabular}
\label{tab:gof}
\end{table} 

\begin{figure}
\hspace*{-0.65in}
\includegraphics[width=1.20\linewidth]{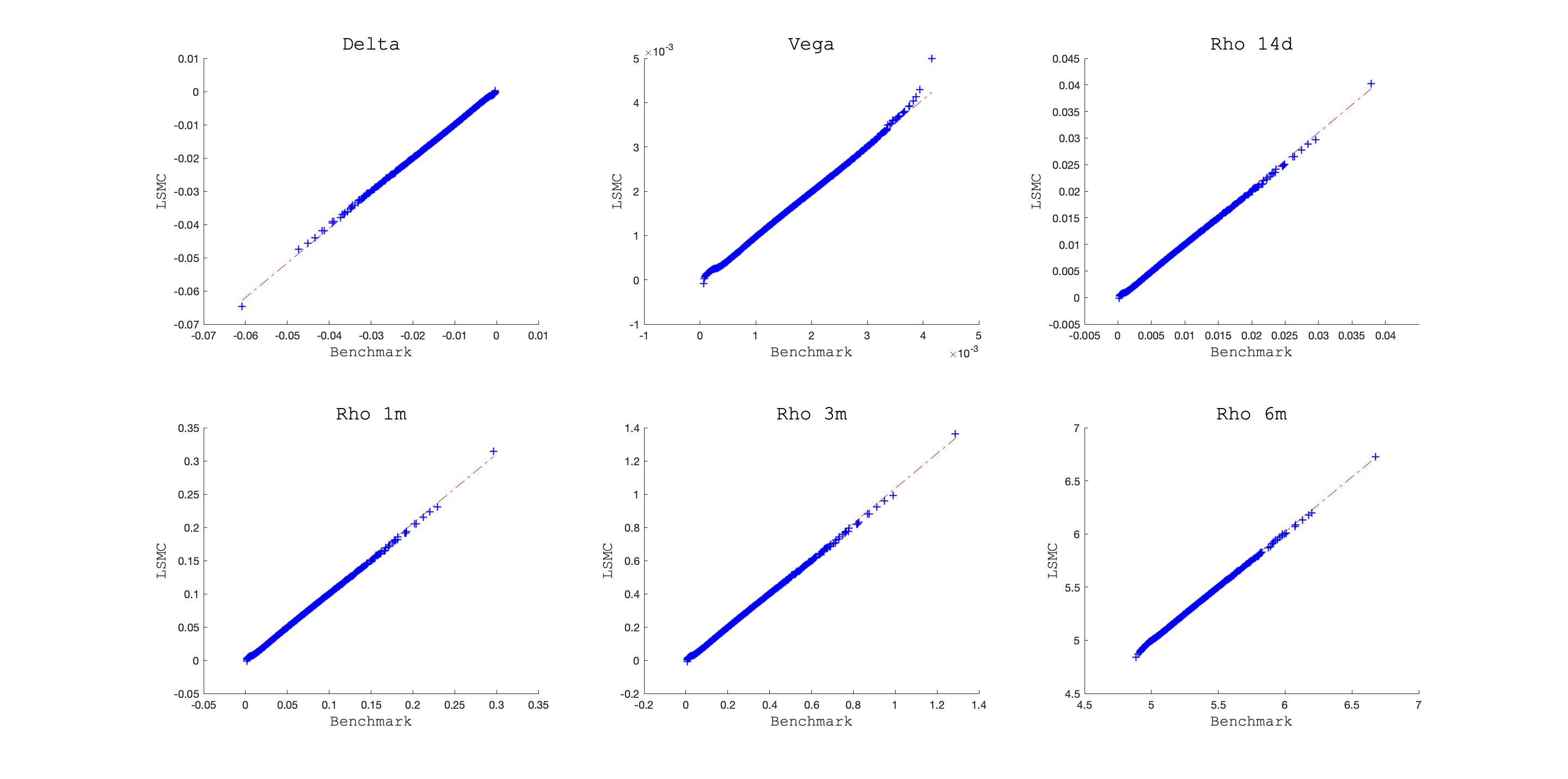}
\captionsetup{justification=centering}
\caption{Arithmetic Asian Option : QQ plot for LSMC estimates using 16 paths.}
\label{fig:asianlsmcqq16}
\end{figure}

Perhaps, a visual graph of the LSMC results is also helpful for us to assess whether the results generated by 
the LSMC plausibly comes from the distribution of the benchmark. This can be achieved using the Quantile-Quantile (QQ) plot. 
A QQ plot is a scatterplot created by plotting two sets of quantiles against one another. If both sets of quantiles come 
from the same distribution, we should see the points forming roughly a straight line. The QQ plot allows us to see at-a-glance if our assumption is plausible, and if not, how the assumption is violated 
and what data points contribute to the violation. Figure \ref{fig:asianlsmcqq16} shows the QQ plot for the
LSMC method using only 16 paths versus the benchmark. Except for the Vega sensitivity, 
the remaining set of results seem to match for most practical purposes. 

A crucial question that we need to answer is that how we select the number of 
inner paths under the LSMC method. It is important to note that our main objective is to not necessarily surpass the accuracy of the FMC method, with say $p$ paths, but to reduce the number of LSMC paths, say $q$,
as much as possible without forsaking the accuracy of the final calculations involving the sensitivities.
In other words, our aim should be to find the most optimal ratio $p/q$ for the targeted application. 
For instance, if a particular sensitivity fails the KS test but its impact on the final calculation 
is limited, then it should still be considered a viable option provided its QQ plot is reasonably a straight line.
We already see from the above example that the ratio $p/q$ depends on the set of sensitivities we need to estimate.
We shall see in the next example that this ratio also depends on the type of payoff. 

Our next example is a Barrier Option. Details of this option are enumerated in
Tables \ref{tab:instr} and \ref{tab:under}. This is a difficult example from
the perspective of the application of the LSMC technique and will truly test its strength. For the Asian option example, we were easily able to obtain very good results with the traditional
polynomial based regression. Based on our previous experiences, we would like to emphasize that such a casual implementation of this method occasionally run into difficulties
when the payoffs are complicated. We would, therefore, not pursue polynomial based
regression any further.\footnote{We are aware of some successful commercial implementations of this approach. Nevertheless, in this paper we mainly advocate the TPS.}
For the case of the Barrier option, we shall equip the LSMC technique with several different
de facto methods used in the risk management industry. In particular, we shall use:
\begin{enumerate} 
\item Smoothing of Barriers to remove numerical instabilities, since the payoff function of Barrier option is discontinuous. \footnote{Numerous research shows that the calculation of the sensitivities based on the finite difference method for discontinuous payoff functions is biased and unstable. \cite{Yuan} provides a good literature summary in this area.}
\item Automatic Adjoint Differentiation (AAD) instead of finite difference method. A very illustrative introduction on the AAD method can be found in \cite{Savine}.
\item Thin Plate Splines (TPS) instead of regular polynomial regression to smooth the Monte Carlo error in the estimates. A brief introduction of TPS was given in Section \ref{sec:LSMC}.
\item Data Thinning to reduce the number of outer scenarios. This is also necessary for efficient implementation of the TPS calibration. This would be achieved using the nearest neighbourhood search algorithm.
\end{enumerate}

We now briefly summarize the above techniques. We start by noting that
the payoff for the Barrier option fails to be Lipschitz continuous. 
If we try to use the finite difference approach to estimate the sensitivities, 
we are bound to run into numerical instabilities. Only by taking prohibitively large number of inner paths are we able to produce some usable estimates.
AAD cannot be used for non Lipschitz continuous payoff function either (c.f. \cite{Giles}.  
\begin{figure}
\centering
\begin{subfigure}{.5\textwidth}
  \centering
  \includegraphics[width=\linewidth]{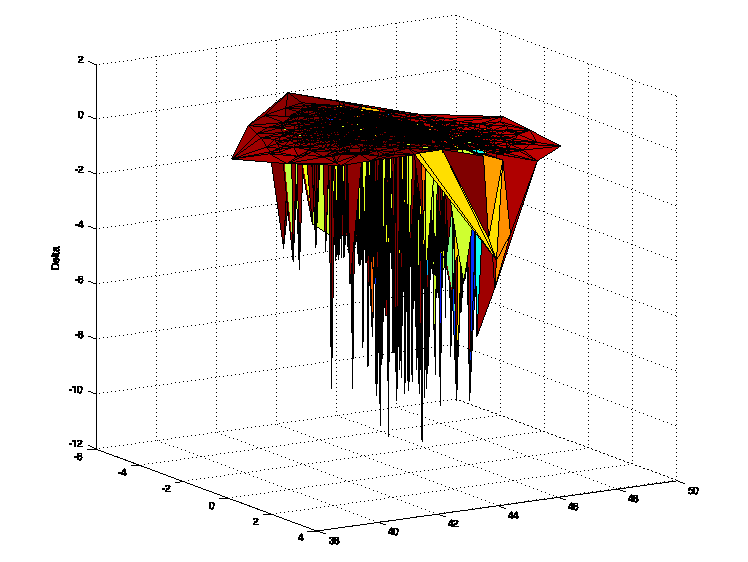}
  \caption{FD method with 4K paths}
  \label{fig:subdelta_4096}
\end{subfigure}%
\begin{subfigure}{.5\textwidth}
  \centering
  \includegraphics[width=\linewidth]{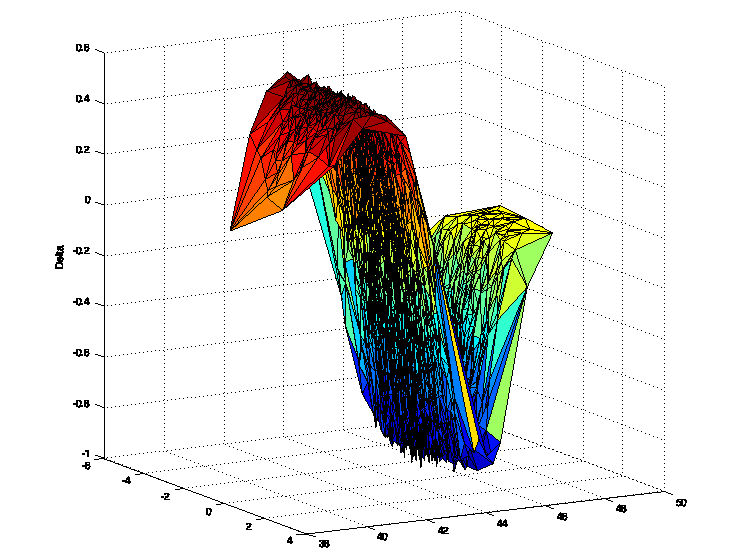}
  \caption{FD method with 16M paths}
  \label{fig:subdelta_16M}
\end{subfigure}
\caption{Instabilities in Delta sensitivity}
\label{fig:bdelta}
\end{figure}
\begin{figure}
\centering
\begin{subfigure}{.5\textwidth}
  \centering
  \includegraphics[width=\linewidth]{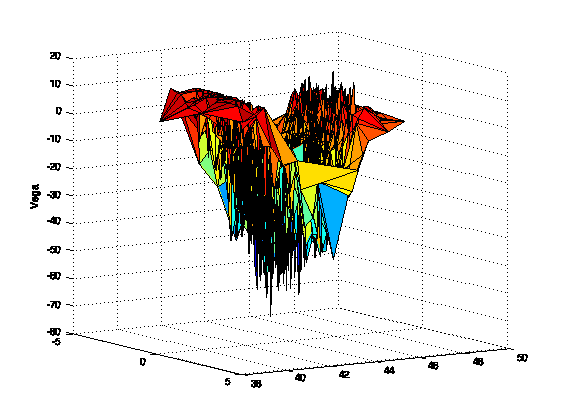}
  \caption{FD method with 4K paths}
  \label{fig:subvega_4096}
\end{subfigure}%
\begin{subfigure}{.5\textwidth}
  \centering
  \includegraphics[width=\linewidth]{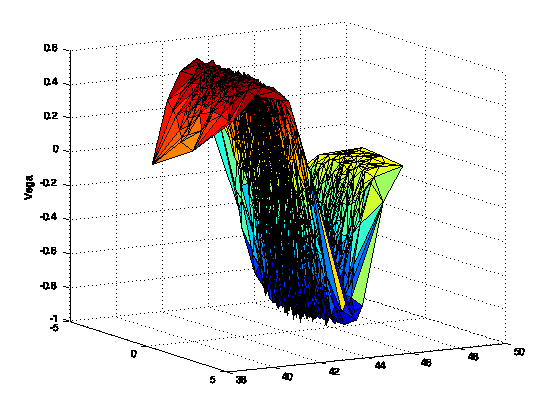}
  \caption{FD method with 16M paths}
  \label{fig:subvega_16M}
\end{subfigure}
\caption{Instabilities in Vega sensitivity}
\label{fig:bvega}
\end{figure}

For our current example, we show the numerical stabilities in the estimate of Delta and Vega sensitivities in Figures \ref{fig:bdelta} and \ref{fig:bvega} respectively. In each of these figures the plot on the left hand side is produced using 4,096 paths, whereas the plot on the right hand side is produced using 16,777,216 paths. These diagrams show the surface plot of the respective sensitivities as a function of stock price and interest rate factor (we are using one factor Hull-White model and keeping the volatility of the stock price constant along the outer scenarios). Clearly, the results obtained using 4,096 paths are unusable for any practical calculations. The results obtained using 16,777,216 paths are somewhat usable, but the cost of computing them is unacceptably high.

This suggests that for all practical purposes, we have to apply some form of smoothing on the Barrier payoff function. We refer to the excellent work by Savine (\cite{Savine}) and Bergomi (\cite{Bergomi}). We follow the former which is based on fuzzy logic to smooth out the ``edge conditions'' of the Barrier option. We use the Logit function to compute the degree of truth. Accordingly, for $L$ Barrier check dates, the degree of truth is calculated using the following formula:
\begin{equation}
\label{eq:dot}
\rm{DoT}(S(1),\ldots,S(L)) = \frac{1}{1+e^{-\epsilon_K(S(L) - K)}} \prod_{l=1}^L \frac{1}{1 + e^{-\epsilon_l (B - S(l))}	},
\end{equation}
where $S(l)$ represent the simulated stock price on the $l$-th fixing date, $K$ denotes the strike and $L$ represents the index of the maturity date.
The various $\epsilon$ factors appearing on the right hand side of Equations (\ref{eq:dot}) are based on the variance of the stock price process. For details, we refer to \cite{Savine} and \cite{Bergomi}. In this paper, however, we take  them as $\epsilon_l = \varepsilon_1 \sigma(l)$ and $\epsilon_K = \varepsilon_2 \sigma(L)$ for some constant $ \varepsilon_1$ and $ \varepsilon_2$. For the purpose of this research, we use an offline grid search algorithm that provides us with suitable values for $\varepsilon_1$ and $\varepsilon_2$ that matches the price and delta sensitivity obtained using 16,777,216 paths. 
 \vskip 0.1in
\begin{rem}{}{}
It should be noted at this point that smoothing of non Lipschitz payoffs is not a direct requirement of the LSMC method. One recalls that LSMC works on raw sensitivity estimates with large Monte Carlo error, but it is not designed to handle unstable estimates.
Even if one were to use FMC (instead of LSMC), one would still need to apply some form of
smoothing - especially considering the fact that adjoint differentiation mandates 
the payoff to be Lipschitz continuous. 
\end{rem}

Once the proper smoothing of the payoff function is in place, we can use the adjoint method to estimate all the required sensitivities in one-pass along with the price. We refer to the work of Giles and Glasserman for details ( \cite{Giles} ). In our case we choose to use the automatic (algorithmic) version of adjoint differentiation approach, known as AAD. For details, readers are suggested to consult \cite{Griewank}. In particular, we choose to use source code transformation (SCT) method to implement the AAD. This is done by using the software Tapenade (c.f. \cite{Hascoet}).

We have already touched briefly on the Thin Plate Splines (TPS) algorithm in the previous section
where we learned that the degree of smoothness is controlled by the parameter $\lambda$. In their paper Wahba and Wendelberger ( \cite{Wahba} ) discuss how to select a proper value of this parameter based on the given data. In this paper, however, we choose a fixed value of lambda and is determined by the equality $n \lambda = 0.5$.\footnote{$n$ is the selected number of outer scenarios obtained from the nearest neighbourhood search algorithm.} We realize that this choice may be sub optimal and, as such, the potential speed ups resulting from LSMC method could be underestimated. Calibration of the optimal value of $\lambda$ is beyond the scope of the current paper.

At this point we would like to point out that the calibration of the TPS is rather costly - especially over a large set of data points. As such, we propose to implement some form of data thinning. Several ideas have been proposed to achieve this feature, almost all of them use the information available in the data of problem ($\bold f$ in equation (\ref{eq:tpscoeff})).  Unfortunately, this is not something we can use under a nested Monte Carlo setting due to its computational cost. We should avoid running a data thinning algorithm at every point in time during a simulation. We would, therefore, prefer that all necessary computations are done at the scenario generation level.
As such, we adopt a simpler approach here. We apply thinning based only on the scenarios -- using the nearest neighbour search (c.f. \cite{Bois} ) algorithm. For example, the space spanned by stock price and interest rate factor would result in a two dimensional nearest neighbour search. 
Points too close to one another are filtered out. 
In our implementation, we specify the number of points to be retained instead of the minimal distance. 
We face a similar situation when using Equation (\ref{eq:tpscoeff}). If we have to execute these calculations at every time step during simulation, it could be expensive. Furthermore, we would have to do this again and again for each option (even when the underlying risk factors are the same). As such, we propose to precompute the matrix $\hat{\bold A}$ and $\hat{\bold B}$ before the start of the simulation. One can also precompute the matrix $\bold A$ incorporating the filtered set of points, so as to obtain the smooth values of the sensitivity estimates for all the outer scenarios.  

As in the case of the Asian option example, the explanatory variables for the Barrier Option comprises of the stock price ($S$) which is modelled by GBM and the PCA factor ($P$) corresponding to the HW1F model. Instead of employing formula (4) to represent the conditional expectation function at some cross section of time, we shall use formula (\ref{eq:tpsexp}) to represent the same in case of the Barrier option example. For numerical experiments, we reduce the number of outer scenarios to be 2,000 by filtering out nearby scenarios using a nearest-neighbourhood search (theoretically, this already should provide us a speedup factor of two), 1,024 inner paths and $n\lambda = 0.5$. 

The benchmark is obtained using 1,048,576 paths nested Monte-Carlo. The results for the Kolmogorov Smirnov statistics (p-values) appear in Table \ref{tab:barriergof}.\footnote{In Table \ref{tab:barriergof}, $R(t,T)$ denotes forward rate from period $t$ to $T$.} Here we also provide the FMC estimate using AAD for 16,384 paths for the sake of comparison.\footnote{16,384 paths were chosen because the FMC estimates fail the Kolmogorov Smirnov tests for all test cases with fewer than 16,384 paths, i.e. for 1024, 2048, 4096 and 8192.}. We observe that the LSMC method with 1024 inner paths and 2000 outer scenarios provide more accurate sensitivity estimations than the FMC (with AAD)  with 16,384 inner paths and 4096 outer scenarios in terms of the KS tests.
 \vskip 0.1in
\begin{rem}{}{}
Our choice of filtering out about half of the outside scenarios and retaining 2000 only is taken rather arbitrarily. One is free to choose not to do any filtering which will increase the cost of using the TPS. Not doing any filtering may also cause issues with smoothing. So some filtering is advisable and in our experiments 2000 seemed to be a good choice.
\end{rem}

\begin{table}[h]
\small
\caption{Barrier Option : Kolmogorov - Smirnov Test}
\centering
\begin{tabular}{!{\VRule[3pt]} l !{\VRule} c !{\VRule} c !{\VRule}
 c !{\VRule} c !{\VRule} c !{\VRule} c !{\VRule} c !{\VRule} c !{\VRule[3pt]} }
\hline
 {} & Delta & Vega & R(0,30) & R(30,60) & R(60,91) & R(91,121) & R(121,151) & R(151,182) \\	
\hline
LSMC & 0.1633 & 0.2730 & 0.1716 & 0.9188 & 0.8361 & 0.9298 & 0.8807 & 0.1267 \\
FMC & 0.1478 & \cellcolor{myGray} 0.0003 & 0.1554 & 0.1203 & 0.1141 & \cellcolor{myGray} 0.0381 & 0.0919 & \cellcolor{myGray} 0.0064 \\
\hline
\end{tabular}
\label{tab:barriergof}
\end{table} 

In implementing the AAD, we chose to compute the sensitivities with respect to the stock price (delta), volatility (vega) and the forward rates. The latter is not what we want. In practice, we need sensitivities with respect to the zero curve (rho). Of course, one is always able to compute zero rate sensitivities from forward rate sensitivities. The results are presented in Table \ref{tab:barriergofzero}. One could also bake in the interpolation algorithm directly in the AAD routine, thereby obtaining the zero rate sensitivities directly.

\begin{table}[h]
\small
\caption{Barrier Option : Kolmogorov - Smirnov Test}
\centering
\begin{tabular}{!{\VRule[3pt]} l !{\VRule} c !{\VRule} c !{\VRule}  c !{\VRule[3pt]} }
\hline
 {} & Rho30 & Rho90 & Rho182  \\	
\hline
LSMC & \cellcolor{myGray}9.58E-08 & 0.1334 & 0.4946 \\
FMC & \cellcolor{myGray} 0.0069  & \cellcolor{myGray} 0.0018 & 0.0778 \\
\hline
\end{tabular}
\label{tab:barriergofzero}
\end{table} 

\begin{figure}
\centering
\begin{minipage}{.5\textwidth}
  \centering
   \includegraphics[width=1\linewidth]{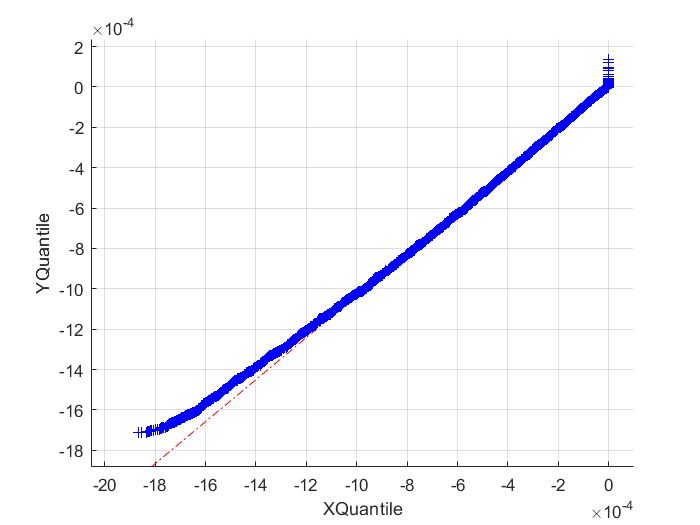}
  \captionof{figure}{QQ Plot : Rho 30day Sensitivity}
  \label{fig:bqqrho}
\end{minipage}%
\begin{minipage}{.5\textwidth}
  \centering
   \includegraphics[width=1\linewidth]{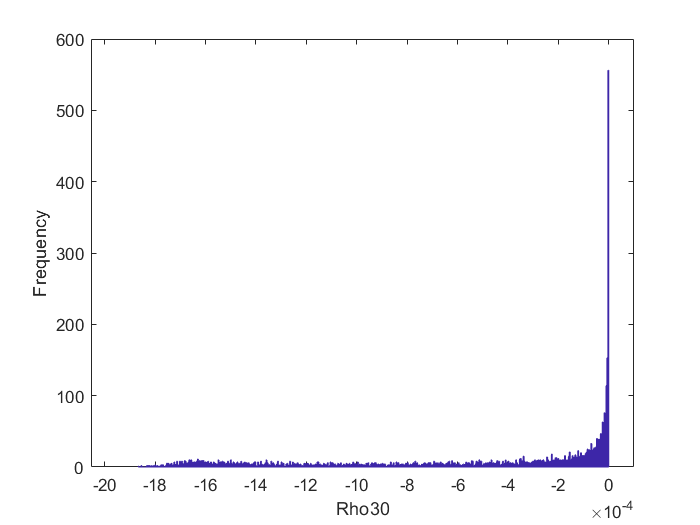}
  \captionof{figure}{Benchmark : Rho 30-day Sensitivity}
  \label{fig:bhistrho30}
\end{minipage}
\end{figure}

From Table \ref{tab:barriergofzero} we observe, as in the case of forward rate sensitivities, that the LSMC method provides better estimations than the FMC method in terms of the KS p-values. We see that the 30 day zero sensitivity fails the 95\% goodness of fit test. We need to investigate what could be wrong with the help of the QQ plot. In Figure \ref{fig:bqqrho} we show the QQ plots for the Rho 30-day sensitivity which does seem to deviate the most near the tails. 
Even though we see deviation on the right tail, it is of little consequence as these numbers are very close to zero and do not have any significant impact on final calculations. LSMC, in general, does have difficulty estimating very small numbers - but from parctical point of view the impact of this is minimal in the final calculation. What should worry us a bit is the deviation on the left end of the tail. That tells us that there could be some instability near the left tail. To see this we inspect the histogram of Rho 30-day sensitivity for the benchmark MC with 1 million inner paths. This appears in Figure \ref{fig:bhistrho30} from which it is clearly noticeable that there is a noisy left tail. We suspect that this is due to the fact that we may not have smoothed the ``barrier edges'' enough.  
From the QQ plot, one can conclude that the results are still practically usable in a SIMM calculation, as aggregation would smooth out small anomalies.

 \vskip 0.1in
\begin{rem}{}{}
Before we conclude this section, we need to address the most important question pertaining
to our proposed algorithm. How does one optimally select the value of smoothing parameter $\lambda$ in the TPS fit. Above, we already mention the work of Wahba and Wendelberger ( \cite{Wahba} ) in this regard. However, in a (near) real time risk management system, any overhead resulting from extra calculations is an undesirable property. On the other hand, we are aware of one commercial installation of the vanilla 
version of LSMC for exposure calculations where the type of polynomials and their respective
powers are configured using an offline learning approach. Based on this observation, we believe that $\lambda$ can be learned offline through a simple forward feed network. This is a topic of future research.
\end{rem}

 \section{SIMM Estimation}\label{sec:SIMMEstimation}
We note that all the underlyings of the derivatives of our test portfolio are common stocks. As such, we only need to compute equity product SIMM values. Meanwhile, due to the fact that our test instruments face only interest rate risk and equity risk, the final SIMM computation formula at the portfolio level can be simplified  from equation (\ref{eq:SIMM}) to the following:

\begin{equation}
\label{eq:simm_port1}
SIMM_{portfolio} =SIMM_{Equity}= \sqrt{  IM_{IR}^{2} +  IM_{ EQ}^2 + 2 \psi_{IR,EQ} IM_{IR}  IM_{EQ}} ,
\end{equation}

with 19\%  correlation between IR and EQ risk class $ \psi_{IR,EQ}$ based on ISDA SIMM methodology. Additionally, because all instruments' underlying are equities rather than interest rates, the initial margin of interest rate risk class can be calculated as: 
\begin{equation}
\label{eq:simm_port2}
IM_{IR} = DeltaMargin_{ IR},
\end{equation}
whereas, the initial margin of equity risk class can be calculated as:
\begin{equation}
\label{eq:simm_port3}
IM_{EQ} = DeltaMargin_{EQ} + VegaMargin_{EQ} + CurvatureMargin_{EQ}.
\end{equation}

Terms appearing on the right hand side of equations (\ref{eq:simm_port2}) and (\ref{eq:simm_port3}) can be computed from the ISDA specification 
entailed in the document \cite{ISDA}.

\begin{figure}
\centering
\captionsetup{justification=centering}
\caption{LSMC Estimation Error For Equity Derivative Portfolio SIMM}
  \includegraphics[width=0.8\linewidth, scale=0.5]{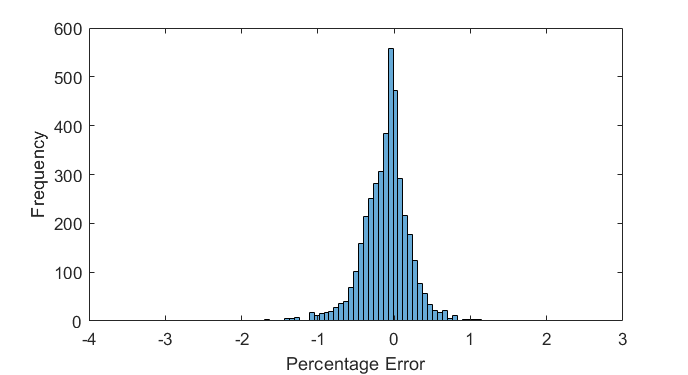}
\label{fig:SIMM_LSMC}
\end{figure}

Based on test results of sensitivities in section \ref{sec:MAIN}, we select 128 inner paths for Asian option and 1024 inner paths for Barrier option under the LSMC framework. The corresponding benchmark initial margin values for Asian option and Barrier option are computed by 1 million inner paths. To make things a little realistic, we append the portfolio with six additional vanilla instruments. These instruments have closed form sensitivity formula, therefore we are able to calculate their IM values directly without using MC simulation. The shaded histogram of relative errors (in \%) of SIMM values at the portfolio level is shown in Figure \ref{fig:SIMM_LSMC}. From this figure we can clearly see that the percentage errors of the LSMC estimation are in the range of [-4\%, 3\%]. Additionally, the p-value of KS test is over 67\%, which indicates we can not reject the null hypothesis that the LSMC estimated SIMM values  come from the same distribution of the benchmark at 95\% confidence level. In order to confirm that the exotic instruments contribute a significant portion in the whole portfolio, we calculate the marginal SIMM value of Asian and Barrier options. The result is 31\%, which demonstrates that our test portfolio is consistent with the market practice. It also demonstrates that the LSMC technique is able to estimate the SIMM values at 95\% confidence level, while significantly reducing the computational cost.

\section{Conclusion}\label{sec:FINAL}
In this paper we demonstrated that the LSMC method described in \cite{lsmc}
for future prices, also works well for obtaining sufficiently accurate and stable estimates of the sensitivities (Greeks). It is able to provide usable values to market practitioners with minimal  computational cost.
This observation has great impact on computing initial margins (SIMM) and capital charges (FRTB).
The LSMC method does have a few limitations. Its ability to capture very small
values is questionable. 
Moreover, LSMC depends on several inputs. For example, the type of polynomial used in regression and their degrees in each direction, the optimal number of inner paths, smoothing parameter $\lambda$ in case of the TPS.  As such, for successful implementation, LSMC should rely on a configuration system to suggest these values. In order to properly build this system, one needs to adopt some form of offline learning and testing. This is a topic of future research. 

\section*{Acknowledgement}
We express special thanks to Shengjie Jin and Alejandra Premat for their constant
support.  We thank Dr. Taehan Bae (University of Regina, Canada) for providing helpful technical hints regarding the lack of convergence of finite difference estimates using the LSMC technique on the bump up and the bump down price estimates.

\end{document}